\documentclass[prb,amsfonts,amssymb,amsmath,showpacs]{revtex4}
\begin{document}
\widetext
\title{Spin-orbit terms in multi-subband electron systems:
A bridge between bulk and two-dimensional Hamiltonians}
\author{K.V.~Kavokin}
\email[E-mail: ]{kidd.orient@mail.ioffe.ru}
\affiliation{A.F.~Ioffe
Physico-Technical Institute, 194021 St.~Petersburg, Russia}
\author{M.E.~Portnoi}
\email[E-mail: ]{m.e.portnoi@exeter.ac.uk}
\affiliation{School of
Physics, University of Exeter, Stocker Road, Exeter EX4 4QL, United
Kingdom}

Published in the special issue of Semiconductors in memory of V. I.
Perel

Phys. Techn. Poluprovodn. {\bf 42}, 1002 (2008) [Semiconductors {\bf
42}, 989 (2008)]


\date{24 June, 2008}

\begin{abstract}
We analyze the spin-orbit terms in multi-subband quasi-two-dimensional
electron systems, and how they descend from the bulk Hamiltonian of the
conduction band. Measurements of spin-orbit terms in one subband alone are
shown to give incomplete information on the spin-orbit Hamiltonian of the
system. They should be complemented by measurements of inter-subband
spin-orbit matrix elements. Tuning electron energy levels with a quantizing
magnetic field is proposed as an experimental approach to this problem.
\end{abstract}
\maketitle

Spin-dependent phenomena in semiconductors was one of the favorite
research themes for Vladimir Idelevich Perel' since the early
1970s.\cite{Perel70th} From the mid-1980s his interests shifted
towards spin-related effects in low-dimensional systems, starting
with optical orientation and polarization properties of hot
photoluminescence in quantum-well structures, which was closely
connected to experiments carried out by the group of D.N.
Mirlin.\cite{Mirlin92}  Some of this work was done together with one
of us (MEP).\cite{PerelPortnoi} In the series of more recent papers,
the transition from the two-dimensional to quasi-three dimensional
case was considered.\cite{quasi3D} Several latest publications of
Vladimir Idelevich were focused on spin-dependent tunneling and the
role played in it by spin-orbit interaction.\cite{PerelLatest} This
has defined the subject choice for our contribution to the special
issue devoted to his memory.

The spin-orbit interaction in semiconductors has been widely discussed
recently in relation to some proposals of spin-electronic and
quantum-computing devices.
It is of considerable physical interest in itself, because due to
strong gradients of atomic potentials within the crystal unit cell the
spin-orbit terms in the effective-mass Hamiltonian are often greatly
enhanced with respect to those of a free electron.\cite{BirPikus}
In addition, the reduced crystal symmetry brings about new spin-orbit
terms unknown for free particles, the so-called Dresselhaus
terms.\cite{Dresselhaus}

In two-dimensional systems, spin-orbit effects are known to be even
stronger than in bulk semiconductors. In particular, in the
effective-mass two-dimensional (2D) Hamiltonian there appear
spin-orbit terms which are linear in the 2D wave vector {\bf k}.
They may exist in structures where the spatial inversion symmetry is
broken. There are so-called bulk inversion asymmetry (BIA) terms,
which appear on averaging the bulk Dresselhaus terms over the
envelope function of the corresponding size-quantization level, and
structure inversion asymmetry (SIA), or Rashba, terms.\cite{Rashba}
The latter are believed to exist in asymmetric quantum wells (QWs);
there is plenty of experimental evidence of their existence in
specific structures, but apparently no agreement has been reached as
to how they descend from the bulk spin-orbit Hamiltonian. Some
authors argue that they are entirely due to interfacial
effects.\cite{Zawadski} The most consistent theoretical treatment of
this problem was carried out by Gerchikov and Subashiev\cite{GS}
(see also a more recent paper by Winkler\cite{Winkler04}). However,
even these papers do not give explicit answers to questions arising
when one attempts to devise experiments aimed at determination of
spin-orbit parameters or to engineer structures with controllable
spin-orbit effects.

In our opinion, for a full understanding of the spin-orbit effects in
nanostructures it is necessary to take into consideration inter-subband
spin-orbit coupling in multi-subband quantum-dimensional structures, which
present a natural bridge between 3D and 2D semiconductor structures.
In this paper we analyze the effect of the spin-orbit interaction
on the energy spectrum of many-subband QWs, and show that applying
a quantizing magnetic fields to such a system presents a way for experimental
determination of all the relevant parameters of the spin-orbit interaction.
Implications for optical and phonon spectroscopy are discussed.

We start with the phenomenological expression for the spin-orbit Hamiltonian
of the conduction band of a compositionally homogeneous bulk semiconductor
in the envelope-function approximation.
It has the following general form:\cite{BirPikus}
\begin{equation}
H_{SO}=({\bf h}_{BIA}\cdot {\bf S)+}a_{V}
{\bf ([k\times \nabla }V{\bf ]\cdot S)},
\label{1}
\end{equation}
where the ``Dresselhaus field'' ${\bf h}_{BIA}$, existing in
non-centrosymmetric crystals, is a pseudovector that is an odd function of
the components of the electron wave vector ${\bf k}$,
and $V$ is the electrostatic potential energy.
The second term in Eq.~\eqref{1} has the same form as the
spin-orbit Hamiltonian of a free electron;
however, the constant $a_{V}$ is not equal to the vacuum spin-orbit
constant $a_{vac}={\hbar ^{2}}/\left(4m_{e}^{2}c^{2}\right)$,
where $m_{e}$ is the bare electron mass.
As electrons in semiconductor crystals are actually subjected
to strong potential gradients within the crystal unit cell,
the spin-orbit interaction is enhanced by a
factor of approximately \thinspace $m_{e}c^{2}/E_{g}$,
where $E_{g}$ is the semiconductor band gap.\cite{AbYass}
This huge enhancement of the spin-orbit interaction has allowed,
in particular, observation of
spin-dependent currents due to anisotropic scattering of electrons by
impurity centers (a solid-state analog of the Mott effect)
in GaAs.\cite{AbYass,AverkDyak,Tkachuk}

In nanostructured semiconductors based on solid solutions like
Ga$_{x}$Al$_{1-x}$As, not only the electrostatic potential,
but also the composition $x$ and, respectively, all the parameters
of the band structure, may depend on the coordinates.
One can therefore introduce a new phenomenological spin-orbit
term proportional to ${\bf \nabla }x$.
Since all the band energies are, to the first approximation,
linear in $x$, we shall write this term for
electrons as $a_{X}{\bf ([k\times \nabla }E_{C}{\bf ]\cdot S)}$.
Here, ${\bf \nabla }E_{C}$ is the ``variable-gap'' field that affects
charge carriers in structures with gradients of composition.\cite{Volkov}

Using the standard 8$\times$8 {\bf{kp}} method of calculation of
spin-orbit splitting, analoguos to that used in
Refs.~\onlinecite{Zawadski,Ohkawa,Lassnig,Bassani}, Gerchikov and
Subashiev\cite{GS} have shown that the spin-orbit term in the
conduction band Hamiltonian can be expressed through a gradient of
an effective ``spin-orbit potential'' $\chi_C$:
\begin{equation}
H_{SO}=({\bf h}_{BIA}{\bf +[k\times \nabla }\chi _C{\bf ])\cdot S},
\label{2}
\end{equation}
where $\chi _{C}$ is a function of the energy positions of the extrema of
the conduction band ($\Gamma _{6}$) and the uppermost valence bands ($\Gamma
_{8}$ and $\Gamma _{7}$):
\begin{equation}
\chi _{C}=\frac{P^{2}\Delta (x)}{3\left[ E-E_{V}(x)\right] \left[
E-E_{V}(x)+\Delta (x)\right]}.
\label{3}
\end{equation}
Here, $P$ is the momentum matrix element between $S$ and $P$ Bloch
states, $E$ is the electron energy, $E_{V}(x)$ is the energy
position of the top of the valence band, and $\Delta (x)$ is the
spin-orbit energy splitting. As noted in Ref.~\onlinecite{GS},
Eq.~\eqref{2} is true even in interface regions, where ${\bf \nabla
}\chi _{C}$ can be expressed in terms of delta functions. In the
following, we shall consider only structures where the energy of
size quantization is much less than the bandgap $E_{g}$. This
condition assumes that variations of both $V$ and $E_{C}$ are much
less than $E_{g}$ over the region where size-quantized wave
functions are mainly concentrated. This class of structures includes
wide quantum wells or heterojunctions with high barriers, but in
this case interfacial regions, where the electron probability
density is small but the gradient of $E_{C}$ is very large, should
be considered separately (see discussion below). For such
structures, one can easily obtain from Eq~(2) of
Ref.~\onlinecite{GS} (compare also Eqs.~(7) and (10) of
Ref.~\onlinecite{Bassani}) the following expressions for $a_{V}$ and
$a_{X}$:
\begin{eqnarray}
a_{V} &=&\frac{\hbar ^{2}\Delta \left( 2E_{g}+\Delta \right) }
{2mE_{g}\left(E_{g}+\Delta \right) \left( 3E_{g}+2\Delta \right)},
\nonumber \\
a_{X} &=&\frac{\hbar ^{2}\Delta \left( 2E_{g}+\Delta \right) }
{2mE_{g}\left(E_{g}+\Delta \right) \left( 3E_{g}+2\Delta \right) }
\left[ \frac{E_{g}^{2}}
{\Delta \left( 2E_{g}+\Delta \right) }\frac{d\Delta}
{dE_{C}}+\frac{dE_{V}}{dE_{C}}\right],
\label{4}
\end{eqnarray}
where $m=3\hbar ^{2}P^{2}E_{g}(E_{g}+\Delta )/2(E_{g}+2\Delta )$ is the
effective mass of the conduction-band electron. For example, using band
parameters and their dependence on composition in Ga$_{1-x}$Al$_{x}$As
from Ref.~\onlinecite{Burstein},
one obtains $a_{V}=4.1\times 10^{-16}$ cm$^{-2}$,
and $a_{X}=-3.1\times 10^{-16}$ cm$^{-2}$ near $x=0$.
For the class of structures we consider, spin-orbit constants, as well
as the effective mass, can be approximately treated as spatially
invariable parameters.

We can now analyze how bulk spin-orbit terms transform as a result of size
quantization in QWs. Due to translational invariance in the
QW plane, the general form of the matrix element of the spin-orbit
Hamiltonian between electron eigenfunctions in the QW is:

\begin{equation}
\left\langle m{\bf k}_{1}\mu \left| \hat{H}_{SO}\right| n{\bf k}_{2}\nu
\right\rangle =({\bf h}_{mn}({\bf k})\cdot {\bf S)}_{\mu \nu }\delta
({\bf k}-{\bf k}_{1})\delta ({\bf k}-{\bf k}_{2}),
\label{5}
\end{equation}
where $m$ and $n$ enumerate size-quantization subbands, ${\bf k}_1$ and
${\bf k}_2$ are electron wave vectors, and $\mu $ and $\nu $ are spin
indices.

Let us first consider the SIA (Rashba) terms. If all the potential gradients
are normal to the QW plane (we denote the corresponding unit vector as
${\bf n}$), the spin-orbit field ${\bf h}_{mn}({\bf k})$ takes the form:
\begin{equation}
{\bf h}_{mn}^{SIA}({\bf k})=A_{mn}^{SIA}[{\bf n}\times {\bf k}],
\label{6}
\end{equation}
where
\begin{eqnarray}
A_{mn}^{SIA}&=&\int\limits_{-\infty }^{+\infty }\Psi _{m}(z)\Psi _{n}(z)
\frac{d\chi _{C}}{dz}dz
\nonumber \\
&=& a_{V}\int\limits_{-\infty }^{+\infty }\Psi _{m}(z)
\Psi_{n}(z)\frac{dV}{dz}dz+a_{X}\int\limits_{-\infty }^{+\infty }\Psi
_{m}(z)\Psi _{n}(z)\frac{dE_{C}}{dz}dz+\sum\limits_{i}I_{mn}^{i}.
\label{7}
\end{eqnarray}
Here, $\Psi_{m}(z)$ and $\Psi_{n}(z)$ are envelope functions of the
$m$th and $n$th subband in the ${\bf z}$-direction, respectively.
The values of $a_{V}$ and $a_{X}$ correspond to the bottom of the
QW. The last term is an interfacial contribution, which arises
because at interfaces sharp gradients of $E_{C}$ coincide with
abrupt changes of $\chi _{C}$; summation is taken over all the
interfaces. Since wave functions are continuous at interfaces, the
overall contribution of interface regions to the matrix element
$A_{mn}^{SIA}$ can be written as $\Psi _{m}(z_{i})\Psi
_{n}(z_{i})(\chi _{C}(z_{i+})- \chi_{C}(z_{i-}))$, where $z_{i+}$
and $z_{i-}$ denote positions immediately to the right and to the
left of the interface, respectively. The term $I_{mn}^{i}$ is
therefore equal to the difference between this expression and the
contribution of the interface to the integral in the second term of
Eq.~\eqref{7}:
\begin{eqnarray}
I_{mn}^{i} &=&\Psi _{m}(z_{i})\Psi _{n}(z_{i})
\left[\chi _{C}(z_{i+})-\chi_{C}(z_{i-})\right]
-\Psi _{m}(z_{i})\Psi_{n}(z_{i})a_{X}
\left[E_{C}(z_{i+})-E_{C}(z_{i-})\right]=
\nonumber \\
&=&\Psi _{m}(z_{i})\Psi _{n}(z_{i})
(\bar{a}_{i}-a_{X})\left[E_{C}(z_{i+})-E_{C}(z_{i-})\right],
\label{8}
\end{eqnarray}
where $\bar{a}_{i}=\left[\chi _{C}(z_{i+})-
\chi_{C}(z_{i-})\right]/\left[E_{C}(z_{i+})-E_{C}(z_{i-})\right]$
has the meaning of an effective constant $a_{X}$, renormalized due
to abrupt changes of the band structure at the interface. For
example, for the ${\rm GaAs/Ga}_{0.6}{\rm Al}_{0.4}{\rm As}$
interface, $\bar{a}_{i}=-2.5\times 10^{-16}$ cm$^{2}$. Since the
bulk constant $a_{X}=-3.1\times 10^{-16}~{\rm cm}^{2}$, the
difference $|\bar{a}_{i}-a_{X}|$ is much smaller than $|a_{X}|$, and
we can conclude that the interfacial corrections are not significant
in the ${\rm GaAs/Ga}_{1-x}{\rm Al}_{x}{\rm As}$ system.

As the QW localization potential for electrons is $U(z)=V(z)+E_{C}(z)$,
Eq.~\eqref{7} can be rewritten in the following way:
\begin{equation}
A_{mn}^{SIA}=a_{X}\int\limits_{-\infty }^{+\infty }
\Psi _{m}(z)\Psi _{n}(z)\frac{dU}{dz}dz
+(a_{V}-a_{X})\int\limits_{-\infty }^{+\infty }
\Psi_{m}(z)\Psi _{n}(z)\frac{dV}{dz}dz
+\sum\limits_{i}I_{mn}^{i}.
\label{9}
\end{equation}
Making use of the fact that $\frac{dU}{dz}=-\dot{p}_{z}$, where
$p_{z}$ is the $z$-component of the electron momentum, we
obtain:\cite{Landau}
\begin{equation}
\int\limits_{-\infty }^{+\infty }\Psi _{m}(z)\Psi
_{n}(z)\frac{dU}{dz}dz =-\frac{i}{\hbar }(E_{n}-E_{m})
\int\limits_{-\infty }^{+\infty }\Psi _{m}(z) \hat{p}_{z}\Psi
_{n}(z)dz=(E_{m}-E_{n})\int\limits_{-\infty }^{+\infty }\Psi
_{m}(z)\frac{d}{dz}\Psi _{n}(z)dz, \label{10}
\end{equation}
where $E_{n}$ and $E_{m}$ are energy levels corresponding to
eigenfunctions $\Psi _{n}(z)$ and $\Psi _{m}(z)$.

Considering the second term in Eq.~\eqref{9},
we recall that ${\cal E}=-(1/e)/(dV/dz)$ is an electric field which
is constant across the quantum well unless there are electric charges
inside the well.
Modern nanostructure technology usually avoids placing impurity atoms
inside the quantum well, so an inhomogeneity of ${\cal E}$ may normally
arise only due to screening by the electron gas in QWs containing
free electrons.
If the concentration of two-dimensional electrons is small, the electric
field inhomogeneity can be neglected. For a constant ${\cal E}$,
$\int\limits_{-\infty }^{+\infty }\Psi _{m}(z)\Psi _{n}(z){\cal E}dz=
{\cal E}\delta_{mn}$.
In this case, we come to the following expression for $A_{mn}^{SIA}$:
\begin{equation}
A_{mn}^{SIA}=a_{X}(E_{m}-E_{n})\int\limits_{-\infty }^{+\infty} \Psi
_{m}(z) \frac{d}{dz}\Psi _{n}(z)dz-(a_{V}-a_{X})e{\cal E}\delta_{mn}
+\sum\limits_{i}I_{mn}^{i}. \label{11}
\end{equation}
The value $A_{nn}^{SIA}=-(a_{V}-a_{X})e{\cal E}$ $+\sum\limits_{i}I_{nn}^{i}$
is the coefficient of the Rashba term in the $n$th subband. One can see that,
contrary to the widespread opinion, the Rashba term is not entirely due to
interfacial effects; in fact, simple estimations show that the interfacial
term is much smaller than the bulk contribution proportional to the electric
field and to the difference of the spin-orbit constants $a_{V}$ and $a_{X}$.
A parabolic quantum well formed by modulation of composition is a good
example illustrating this property of SIA spin-orbit terms.
There are no interfaces in the parabolic quantum well.
Moreover, applying an electric field in this case does not change the shape
of the localizing potential, which remains symmetric:
\begin{equation}
U(z)=\frac{m\omega ^{2}z^{2}}{2}-{\cal E}ez=\frac{m\omega ^{2}}{2}\left(
z-z_{0}\right) ^{2}-\frac{m\omega ^{2}z_{0}^{2}}{2},
\label{12}
\end{equation}
where $z_{0}={\cal E}e/m\omega ^{2}$.
However, according to Eq.~\eqref{11}, there
exist both intra- and intersubband SIA terms:
\begin{eqnarray}
A_{n-1,n}^{SIA} &=&A_{n,n-1}^{SIA}= a_{X}\sqrt{\frac{m\hbar
\omega^{3} n}{2}},
\nonumber \\
A_{n,n}^{SIA} &=&-(a_{V}-a_{X})e{\cal E}.
\label{13}
\end{eqnarray}
Intra-subband SIA terms are zero in the absence of electric fields, while
inter-subband SIA matrix elements are always present and do not depend on
the electric field. Remarkably, bandgap gradients do not contribute to the
Rashba term: one needs to apply an electric field to the structure to
produce it. These gradients are needed, however, to confine the electron
in the ${\bf z}$-direction; if the electron was confined by a purely
electrostatic potential, Rashba terms would also be absent.
This fact has been noted by Gerchikov and Subashiev.\cite{GS}

Another instructive example is a symmetric rectangular quantum well in an
electric field. Since in this case all the changes of composition $x$ are
concentrated at interfaces, the interfacial effects can be exactly accounted
for by replacing $a_{X}$ with $\bar{a}_{i}$ in Eq.~\eqref{11};
this substitution eliminates the interfacial term.
One can see that in ${\rm GaAs/Ga_{0.6}Al_{0.4}As}$
structures the interfacial correction amounts to less than 10\% of the
intrasubband Rashba term, and about 20\% of the intersubband SIA
matrix element.

For a single heterojunction and other structures where the electric field is
created by separated charges of impurities in depleted doped regions and the
2D electron gas itself, Eq.\eqref{11} should be modified by taking into
account the electric field inhomogeneity. In this case, the electric field
$\mathcal{E}$ in the Rashba term should be replaced with its average value
(integrated with the squared wave function of the corresponding subband).
The field inhomogeneity will also contribute to the inter-subband term.
The correponding corrections can be evaluated using the Poisson equation.
For example, the leading correction to the first term in Eq.\eqref{11}
is: $\tilde{A}_{mn}^{SIA}=(a_{V}-a_{X})\,(4\pi n_{s}e/\varepsilon)
\int\limits_{-\infty }^{+\infty }\Psi_{m}(z)\Psi_{n}(z)
\int\limits_{-\infty }^{z}\Psi _{0}^{2}(z^{\prime })\,dz^{\prime}\,dz$,
where $m\neq n$, $n_{s}$ is the sheet concentration of electrons,
$e$ is the absolute value of the electron charge, $\varepsilon$ is the
dielectric constant, and $\Psi _{0}^{2}$ is the squared
ground-state wave function (here we assume that most
electrons are in the lowest size-quantization subband; otherwise
summation over all occupied subbands should be performed).

BIA (Dresselhaus) terms in bulk zinc-blende semiconductors have the form
$\hat H_{BIA}={\bf h}^{bulk}\left( {\bf k}\right) \cdot {\bf S}$, with
\begin{equation}
h_{x}^{bulk}=\alpha \hbar ^{3}\left( m_{e}\sqrt{2m_{e}E_{g}}\right)
^{-1}k_{x}\left( k_{y}^{2}-k_{z}^{2}\right),
\label{14}
\end{equation}
where $m_{e}$ is the effective mass of the electron, $E_{g}$ is the band
gap, and $k_{x}$, $k_{y}$, $k_{z}$ are components of the wave vector along
the cubic axes [100], [010], and [001] respectively.
The ${\bf y}$- and $\bf {z}$-components of ${\bf h}^{bulk}$ are obtained
from Eq.~\eqref{14} by permutation of indices.

The expression for the BIA matrix elements between 2D subbands depends on
the QW orientation with respect to crystal axes.
For the most common case of a [100] quantum well
\begin{eqnarray}
{\bf h}_{mn}^{{\rm BIA}} &=& A_{mn}^{{\rm BIA}}\left( k_{x}{\bf
e}_{x}-k_{y}{\bf e}_{y}\right) + i\frac{\alpha \hbar
^{3}}{m\sqrt{2mE_{g}}}
\nonumber \\
&\times& \int\limits_{-\infty }^{+\infty }\Psi
_{m}(z)\frac{d}{dz}\Psi _{n}(z)dz\left( k_{x}^{2}-k_{y}^{2}\right)
{\bf e}_{z}, \label{15}
\end{eqnarray}
where
\begin{equation}
A_{mn}^{{\rm BIA}}=\frac{\alpha \hbar ^{3}}{m\sqrt{2mE_{g}}}%
\int\limits_{-\infty }^{+\infty }\Psi _{m}(z)\frac{d^{2}}{dz^{2}}\Psi
_{n}(z)dz.
\label{16}
\end{equation}
A comparison of Eqs.~\eqref{11} and \eqref{15} shows
that BIA and SIA terms in quantum wells are related to the
corresponding bulk Hamiltonians in different ways.
Intra-subband BIA\ terms are determined by a unique constant $\alpha $,
the same as in bulk BIA\ terms.
This is confirmed experimentally: the values of $\alpha $
measured from spin relaxation in bulk GaAs,\cite{OO} and from
spin-flip Raman scattering in GaAs/AlGaAs quantum wells,\cite{Jusserand}
are indeed very close (0.07 and 0.065, respectively).
On the contrary, Rashba terms depend on a specific combination of
constants ($a_{V}-a_{X}$)
describing different contributions from gradients of electrostatic and
crystal potentials, while, for instance, only $a_{V}$ contributes to
spin-dependent scattering from charged impurities in the bulk. These
constants cannot be determined separately, only by measuring spin-orbit
splitting in size-quantization subbands. These measurements do not,
therefore, give complete information about the spin-orbit Hamiltonian of
the conduction band.

Experimental determination of inter-subband spin-flip matrix elements, which
would be very helpful in this view, is hindered by the fact that they are
typically much smaller than the energy of size quantization, and all the
observable effects of these matrix elements are consequently suppressed. We
should like to note that this difficulty can, in principle, be overcome by
tuning energy levels of quasi-2D electrons with a strong magnetic field.
When a magnetic field is applied normal to the QW plane, the in-plane motion
of electrons is also quantized, and the continuum energy spectrum of each 2D
subband is transformed into discrete Landau levels (LLs). Landau levels
belonging to different subbands $n$ and $m$ can be tuned in resonance by
choosing the magnetic field so that $\left| E_{n}-E_{m}\right| =l\hbar
\omega _{c}$, where $l$ is an integer, and $\omega _{c}$ is the cyclotron
frequency. Under these conditions, the energy spectrum and the structure of
wave functions of the four-level subsystem formed by Zeeman-split spin
components of the two LLs from different subbands are expected to be strongly
affected by the spin-orbit interaction.

Using the Landau gauge ($A_{x}=-Hy$; $A_{y}=A_{z}=0$) and
Eqs.~\eqref{5}, \eqref{6}, \eqref{11}, \eqref{15} and \eqref{16}, we
obtain the following expressions for matrix elements between
electron states differing by subband number, LL number, and spin
projection on the normal to the structure plane:
\begin{equation}
\left\langle \eta ,n,k_{x},-1/2\left| H_{SO}\right|
\mu,n-1,k_{x},+1/2\right\rangle =iA_{\eta \mu }^{SIA}
\frac{1}{l_{H}}\sqrt{2n}
\label{17},
\end{equation}
\begin{equation}
\left\langle \eta ,n-1,k_{x},+1/2\left| H_{SO}\right|
\mu,n,k_{x},-1/2\right\rangle =-iA_{\eta \mu }^{SIA}
\frac{1}{l_{H}}\sqrt{2n},
\label{18}
\end{equation}
\begin{equation}
\left\langle \eta ,n,k_{x},+1/2\left| H_{SO}\right|
\mu,n-1,k_{x},-1/2\right\rangle =A_{\eta \mu }^{BIA}
\frac{1}{l_{H}}\sqrt{2n},
\label{19}
\end{equation}
\begin{equation}
\left\langle \eta ,n-1,k_{x},-1/2\left| H_{SO}\right|
\mu,n,k_{x},+1/2\right\rangle =A_{\eta \mu }^{BIA}
\frac{1}{l_{H}}\sqrt{2n},
\label{20}
\end{equation}
where indices $\eta $ and $\mu $ enumerate subbands, $n$ is the LL
number, and $l_{H}$ is the magnetic length. Remarkably, states from
different subbands and LLs are coupled by either BIA or SIA terms,
but not by the two types of spin-orbit terms at the same time:
Eqs.~\eqref{17} and \eqref{18} couple states $\left\langle
n-1,+1/2\right|$ and $\left\langle n,-1/2\right| $, while
Eqs.~\eqref{19} and \eqref{20} couple states $\left\langle
n-1,-1/2\right|$ and $\left\langle n,+1/2\right|$. Therefore, by
tuning a specific pair of levels into resonance one can prepare an
effective two-level system whose spectrum is determined by either
the BIA or SIA term. As Eqs.~\eqref{17}-\eqref{20} are based on
general symmetry properties of Rashba and Dresselhaus terms rather
than on any specific method of derivation of corresponding
constants, they can be used for experimental determination of
spin-orbit parameters.

The exact form of this spectrum and its experimental manifestations
depend on the electron concentration in the quasi-2D system. If the
concentration is low and many-body effects are absent, the
inter-subband spin-orbit coupling results in anticrossing of the two
levels. The resulting energy gaps, equal to corresponding
inter-subband matrix elements (Eqs.~\eqref{17}-\eqref{20}) can be
measured by high-resolution spectroscopic techniques (optical or
spin/cyclotron resonance spectroscopy), as has been proposed for
one-subband systems in tilted magnetic fields.\cite{Falko} The
admixture of wave functions with opposite spin near the anticrossing
can, in principle, be detected using optical polarization
spectroscopy. If all the electron states below the two selected
levels are filled, many-body effects have been shown to strongly
affect the spectrum of electrons. In particular, they result in the
opening of gaps in the energy spectrum even in the absence of
spin-orbit interaction.\cite{AP} It has been proposed to employ
non-equilibrium phonons to probe this system of strongly coupled
electrons.\cite{AP} In this type of experiment, the spin-orbit
inter-subband coupling would manifest itself by allowing
phonon-assisted transitions between levels with different
projections of the electron spin, which are otherwise
forbidden.\cite{AP,GP}.

In conclusion, we have analyzed the transition from bulk to two-dimensional
behavior of the spin-orbit interaction in multi-subband
quasi-two-dimensional heterostructures. Intra- and inter-subband matrix
elements of the spin-orbit Hamiltonian are derived, and the role of
interfacial effects is estimated. The two types of spin-orbit terms
(Rashba and Dresselhaus) can be distinguished by measuring the
inter-subband coupling of Landau levels in quantizing magnetic fields.

We are grateful to the Editors for inviting us to submit a paper to
the special issue devoted to the memory of V.I. Perel'. His work and
unique personality strongly influenced our research and lit up our
lives.

\bibliographystyle{apsrev}

\begin{thebibliography}{20}

\bibitem{Perel70th} M.I. Dyakonov and V.I. Perel, JETP Lett. {\bf 13},
144 (1971);  M.I. Dyakonov and V.I. Perel,  JETP Lett. {\bf 13}, 467
(1971);  M.I. Dyakonov and V.I. Perel, Phys. Lett. A {\bf 35}, 459
(1971); M.I. Dyakonov and V.I. Perel,  Sov. Phys. JETP {\bf 33},
1053 (1971); for the review of early work see {\it Optical
Orientation}, edited by F. Meier and B.P. Zakharchenya
(North-Holland, Amsterdam, 1984; Nauka, Leningrad, 1989).

\bibitem{Mirlin92} D.N. Mirlin and V.I. Perel, Semicond. Sc.
Technol. {\bf 7}, 1221 (1992).

\bibitem{PerelPortnoi} I.A. Merkulov, V.I. Perel and M.E. Portnoi,
Zh. Exp. Teor. Fiz. {\bf 99}, 1202, 1991 [Sov. Phys. JETP {\bf 72},
669 (1991)]; I.A. Merkulov, V.I. Perel and M.E. Portnoi, Supperlatt.
Microstruct. {\bf 10}, 371 (1991), V.I. Perel' and M.E. Portnoi,
Fiz. Tekh. Poluprovodn. {\bf 26}, 2112 (1992) [Sov. Phys. Semicond.
{\bf 26}, 1185 (1992)].

\bibitem{quasi3D} V.F. Sapega, V.I. Perel, A.Yu. Dobin, D.N. Mirlin,
I.A. Akimov, T. Ruf, M. Cardona, K. Eberl. Phys. Rev. B {\bf 56},
6871 (1997); V.F. Sapega, V.I. Perel', D.N. Mirlin, I.A. Akimov, T.
Ruf, M. Cardona, W. Winter, and K. Eberl, Fiz. Tekh. Poluprovodn.
{\bf 33}, 738 (1999) [Semiconductors {\bf 33}, 681 (1999)]; I.A.
Akimov, D.N. Mirlin, V.I. Perel', and V.F. Sapega, Fiz. Tekh.
Poluprovodn. {\bf 35}, 758 (2001) [Semiconductors {\bf 35}, 727
(2001)].

\bibitem{PerelLatest} V.I. Perel', S.A. Tarasenko, I.N. Yassievich, S.D. Ganichev,
V.V. Bel'kov, and W. Prettl, Phys. Rev. B {\bf 67}, 201304 (2003);
I.N. Yassievich and V.I. Perel, Physica B {\bf 340}, 496 (2003);
S.A. Tarasenko, V.I. Perel', and I.N. Yassievich, Phys. Rev. Lett.
{\bf 93}, 056601 (2004).

\bibitem{BirPikus} G.L. Bir and G.E. Pikus, {\it Symmetry and Strain-induced
Effects in Semiconductors} (Wiley, New York, 1974), Secs. 25 and 26.

\bibitem{Dresselhaus}  G. Dresselhaus, Phys. Rev. {\bf 100}, 580 (1955).

\bibitem{Rashba}  E.I. Rashba, Fiz. Tv. Tela (Leningrad) {\bf 2}, 1224 (1960)
[Sov. Phys. Solid State {\bf 2}, 1109 (1960)]; Yu.A. Bychkov and E.I. Rashba,
Usp. Fiz. Nauk {\bf 146}, 531 (1985) [Sov. Phys. Usp. {\bf 28}, 632 (1985)].

\bibitem{Zawadski}  P. Pfeffer and W. Zawadzki, Phys. Rev. B {\bf 59}, R5312
(1999).

\bibitem{GS}  L.G. Gerchikov and A. V. Subashiev,
Fiz. Tekh. Poluprovodn. {\bf 26}, 131 (1992)
[Sov. Phys. Semicond. {\bf 26}, 73 (1992)].

\bibitem{Winkler04} R. Winkler, Physica E {\bf 22}, 450 (2004).

\bibitem{AbYass}  V.N. Abakumov, V.V. Akulinichev, and
I.N. Yassievich, Fiz. Tekh. Poluprovodn. {\bf 9}, 936 (1975)
[Sov. Phys. Semicond. {\bf 9}, 612 (1975)].

\bibitem{AverkDyak}  N.S. Averkiev and M.I. D'yakonov,
Fiz. Tekh. Poluprovodn. {\bf 17}, 629 (1983)
[Sov. Phys. Semicond. {\bf 17}, 393 (1983)].

\bibitem{Tkachuk}  A.A. Bakun, B.P. Zakharchenya, A.A. Rogachev, M.N. Tkachuk,
and V.G. Fleisher, Pis'ma Zh. Eksp. Teor. Fiz. {\bf 40}, 464 (1984)
[JETP Lett. {\bf 40}, 1293 (1984)].

\bibitem{Volkov}  A.S. Volkov, A.L. Lipko, Sh.M. Meretliev,
and B.V. Tsarenkov,
Pis'ma Zh. Eksp. Teor. Fiz. {\bf 41}, 458 (1985)
[JETP Lett. {\bf 41}, 557 (1985)].

\bibitem{Ohkawa} F.J. Ohkawa and Y. Uemura, J. Phys. Soc. Jpn. {\bf 37},
1325 (1974).

\bibitem{Lassnig} R. Lassnig, Phys. Rev. B {\bf 31}, 8076 (1985).

\bibitem{Bassani}  E.A. de Andrada e Silva, G.C. La Rocca, and F. Bassani,
Phys. Rev. B {\bf 50}, 8523 (1994).

\bibitem{Burstein} {\it Physics of Group IV Elements and III-V Compounds},
edited by O. Madelung, Landolt-B\"{o}rnstein: Numerical Data and Functional
Relationships in Science and Technology
(Springer-Verlag, Berlin, 1982), Vol. 17, Subvolume A.

\bibitem{Landau} L.D. Landau and E.M. Lifshitz, {\it Quantum Mechanics}
(Pergamon Press, New York, 1977).

\bibitem{OO}  {\it Optical Orientation}, edited by F. Meier and
B.P. Zakharchenya, Modern Problems in Condensed Matter Sciences
(North-Holland, Amsterdam, 1984), Vol. 8.

\bibitem{Jusserand}  B. Jusserand, D. Richards, G. Allan, C. Priester,
and B. Etienne, Phys. Rev. B {\bf 51}, R4707 (1995).

\bibitem{Falko}  V.I. Falko, Phys. Rev. B {\bf 46}, R4320 (1992).

\bibitem{AP} V.M. Apalkov and M.E. Portnoi, Phys. Rev. B {\bf 65},
125310 (2002); V.M. Apalkov and M.E. Portnoi, Physica E {\bf 15},
202 (2002).

\bibitem{GP} V.N. Golovach and M.E. Portnoi, Phys. Rev. B {\bf 74},
085321 (2006).

\end{thebibliography}

\end{document}